\documentclass[nofootinbib,prd,superscriptaddress,twocolumn,reprint]{revtex4-1}
\usepackage{graphicx}
\usepackage{subfigure}
\usepackage{amsmath, amssymb}
\usepackage{slashed}

\pdfoutput=1

\newcommand{\lsim}{\lesssim}

\newcommand{\eq}[1]{Eq.~(\ref{#1})}
\newcommand{\beq}{\begin{equation}}
\newcommand{\eeq}{\end{equation}}
\newcommand{\bea}{\begin{eqnarray}}
\newcommand{\eea}{\end{eqnarray}}
\newcommand{\nn}{\nonumber}

\newcommand{\eps}{\varepsilon}

\newcommand{\mzd}{m_{Z_d}}

\newcommand{\sstwmZ}{\sin^2\theta_W(m_Z)_{\rm \overline{MS}}}
\newcommand{\sstwQ}{\sin^2\theta_W(Q^2)}

\newcommand{\br}{{\rm BR}}

\newcommand{\gev}{{\rm GeV}}
\newcommand{\mev}{{\rm MeV}}

\begin{document}
\title{\boldmath
Low $Q^2$ Weak Mixing Angle Measurements and Rare Higgs Decays}
\author{Hooman Davoudiasl}
\affiliation{
Department of Physics, Brookhaven National Laboratory, Upton, New York 11973, USA}
\author{Hye-Sung Lee}
\affiliation{CERN, Theory Division, CH-1211 Geneva 23, Switzerland}
\author{William J. Marciano}
\affiliation{
Department of Physics, Brookhaven National Laboratory, Upton, New York 11973, USA}

\preprint{CERN-PH-TH-2015-151}

\begin{abstract}

A weighted average weak mixing angle $\theta_W$ derived from relatively low
$Q^2$ experiments is compared with the Standard Model prediction obtained from precision measurements.
The approximate 1.8 sigma discrepancy is fit with an intermediate mass ($\sim 10-35$~GeV)
``dark" $Z$ boson $Z_d$, corresponding to a $U(1)_d$ gauge symmetry of hidden dark matter, which
couples to our world via kinetic and $Z$-$Z_d$ mass mixing.
Constraints on such a scenario are obtained from precision electroweak bounds and searches for the rare Higgs decays $H \to Z Z_d \to $ 4 charged leptons
at the LHC.  The sensitivity of future anticipated low $Q^2$ measurements of $\sin^2\theta_W(Q^2)$
to intermediate mass $Z_d$ is also illustrated.
This dark $Z$ scenario can provide interesting concomitant signals in low energy parity violating measurements and rare Higgs decays at the LHC, over the next few years. 


\end{abstract}
\maketitle

Discovery of what appears to be a fundamental Higgs scalar \cite{Aad:2012tfa,Chatrchyan:2012ufa} completes the basic Standard Model (SM) particle
spectrum.  In addition, comparing precision fine structure constant $\alpha$, Fermi constant $G_F$, and $Z$ boson mass ($m_Z$) values at the quantum loop level,
employing the Higgs mass $m_H = 125$~GeV and top quark mass $m_t = 173.3(8)$~GeV gives the indirect
SM weak mixing angle prediction \cite{Kumar:2013yoa,Agashe:2014kda}
\beq
\sstwmZ = 0.23124(12)\quad \text{SM prediction},
\label{sin2-SM}
\eeq
where the modified minimal subtraction ($\overline{\text{MS}}$) definition at scale $\mu=m_Z$ for
the renormalized weak mixing angle $\theta_W$ has been employed \cite{runningWeinberg}.  The existing error in \eq{sin2-SM} stems from $m_t$, higher order loops (that overall double the error),
and hadronic uncertainties, all of which are expected to be further reduced.
That prediction agrees remarkably well with the average value \cite{Kumar:2013yoa} of the more direct $Z$ pole measurements \cite{Abe:2000dq,ALEPH:2010aa}
\beq
\sstwmZ = 0.23125(16)\quad \text{$Z$ pole average}.
\label{sin2-Zpole}
\eeq
A comparison of these distinct precision methods severely constrains ``new physics" extensions of the SM \cite{Kumar:2013yoa}.

\begin{figure}[t]
\includegraphics[width=0.48\textwidth]{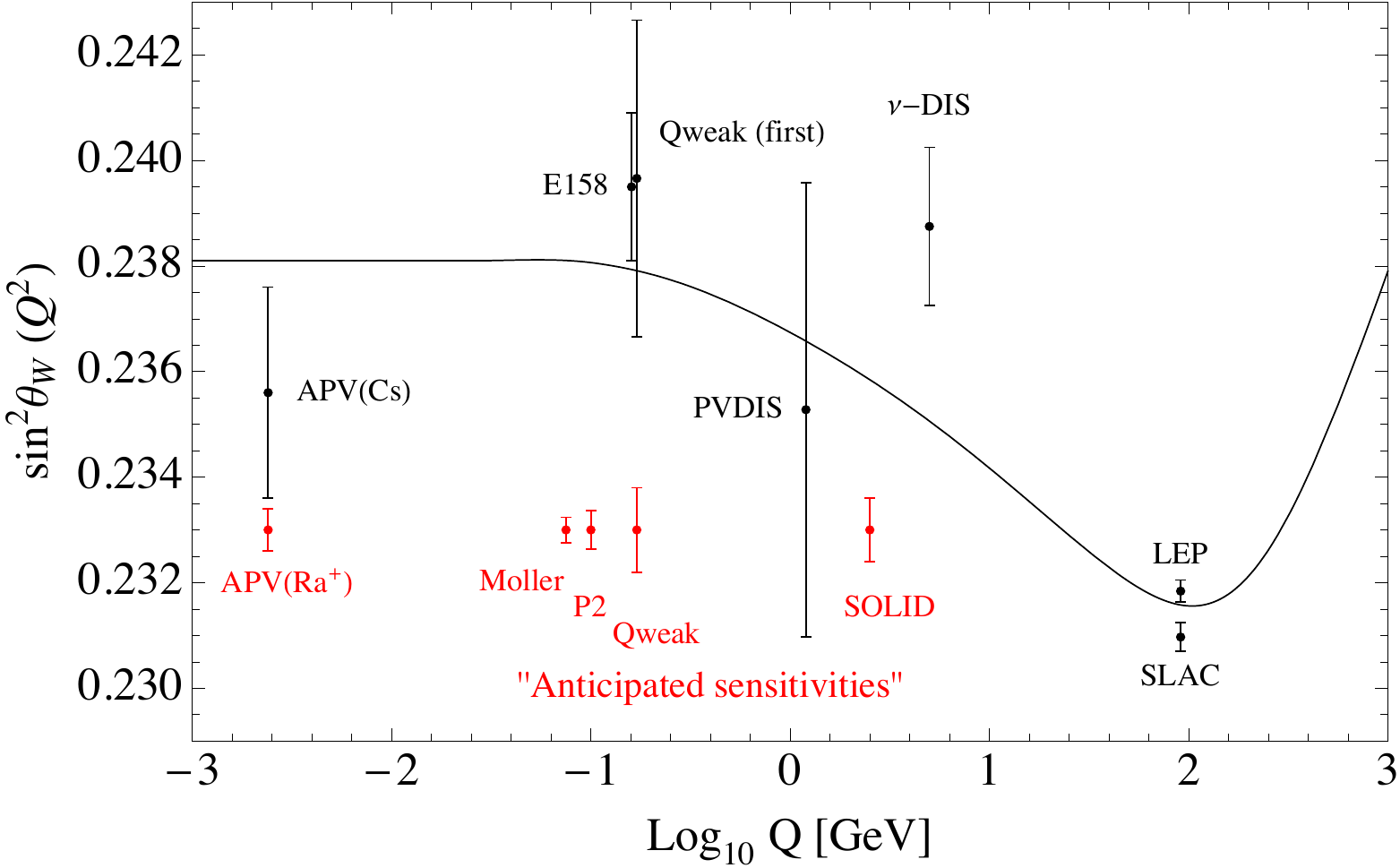}
\caption{Current measurements of the weak mixing angle at various $Q$ \cite{Gilbert:1986ki,Wood:1997zq,Bennett:1999pd,Anthony:2005pm,Zeller:2001hh,Wang:2014bba,Androic:2013rhu,Abe:2000dq,ALEPH:2010aa} and future prospects \cite{NunezPortela2014,Jungmann:2014kia,Mammei:2012ph,MESAP2,Reimer:2012uj}.
The black curve represents the expected SM prediction for the running of $\sin^2\theta_W$ with $Q$ \cite{runningWeinberg}. Current measurements are given as black points with existing error bars.  The red ``Anticipated sensitivities'' are meant only to illustrate the possible uncertainties potentially obtainable from experiments under analysis and proposed.}
\label{fig:SMrun}
\end{figure}

In contrast, low $Q^2$ determinations of the weak mixing angle (for a review, see Ref.~\cite{Kumar:2013yoa}) currently allow considerable room for certain types of new physics, particularly $Z'$ bosons (for earlier work along these lines, see for example Refs.~\cite{Marciano:1990dp,Davidson:2001ji,Boehm:2004uq,Langacker:2008yv}).
Indeed, the 3 most precise measurements at lower $Q^2\ll m_Z^2$ extrapolated, for comparison, to an $\overline{\text{MS}}$
scale $\mu = m_Z$ give a somewhat disparate range of values \cite{Kumar:2013yoa}
\beq
\sstwmZ = 0.2283(20)\quad \text{APV} \label{sin2-APV},
\eeq
\beq
\sstwmZ = 0.2329(13)\quad \text{Moller E158},
\label{sin2-E158}
\eeq
\beq
\sstwmZ = 0.2356(16)\quad \text{NuTeV}
\label{sin2-NuTeV}
\eeq
from the measurements in Cs atomic parity violation (APV) at $\left< Q \right> = 2.4~\mev$  \cite{APVtheory,Gilbert:1986ki,Wood:1997zq,Bennett:1999pd}, SLAC Moller scattering experiment E158 at $\left< Q \right> = 160~\mev$  \cite{Anthony:2005pm}, and Fermilab neutrino deep inelastic scattering (DIS) experiment NuTeV at $\left< Q \right> \approx 5~\gev$ \cite{Zeller:2001hh}.

These measurements are illustrated in Fig.~\ref{fig:SMrun}, after evolving back to their
experimental $Q$ values.  There, we also show other less precise determinations
of $\sin^2\theta_W(Q^2)$ (JLAB Qweak first result \cite{Androic:2013rhu} and JLAB PVDIS \cite{Wang:2014bba}) as well as the very accurate $Z$ pole values \cite{Abe:2000dq,ALEPH:2010aa}, future sensitivities (Ra$^+$ APV \cite{NunezPortela2014,Jungmann:2014kia}, JLAB Moller \cite{Mammei:2012ph}, MESA P2 \cite{MESAP2}, JLAB DIS experiment SOLID \cite{Reimer:2012uj}),
and the predicted SM running curve for comparison.
Note that the Qweak result in our figures corresponds to only about 4\% of their total collected data.
Their statistical uncertainty may be significantly reduced in the near future making them the expectedly best low $Q^2$ determination.
We return to this point later.
Note, also, that the factor of 5 improvement envisioned for APV using single ionized Ra$^+$ trapped atoms as originally suggested in Ref.~\cite{Portela:2013twa}, although extremely well motivated, is still in a development stage \cite{Fortson:1993zz}.
The potential polarized electron scattering asymmetry improvements are currently on a more definite footing.

The weighted average
from Eqs.~(\ref{sin2-APV})-(\ref{sin2-NuTeV})
\beq
\sstwmZ = 0.2328(9)\quad \text{low $Q^2$ average}
\label{sin2-lowQ2}
\eeq
is roughly 1.8 sigma higher than the SM prediction in \eq{sin2-SM}
\beq
\Delta \sin^2\theta_W \simeq 0.0016(9)
\label{sin2-dev}
\eeq
and gives about the same deviation relative to \eq{sin2-Zpole}.

Of course, there are still outstanding issues regarding atomic parity violation theory \cite{Dzuba:2002kx,Porsev:2010de,Dzuba:2012kx} that warrant further scrutiny.
In addition, NuTeV hadronic effects \cite{Bentz:2009yy} and radiative corrections \cite{Marciano:1980pb,Sirlin:1981yz} could shift the average somewhat \cite{Kumar:2013yoa}.
However, here, we take the current average
in \eq{sin2-lowQ2} at face value and examine its consequences for an intermediate mass dark $Z$ ($Z_d$) with
$\mzd \sim 10-35$~GeV (the intermediate mass range bounded from below by 
the onset of severe constraints from low energy measurements and from above by $m_H - m_Z$) and coupling to the SM particles via kinetic and $Z$-$Z_d$ mass matrix mixing.
Although the current
1.8 sigma discrepancy is far from compelling evidence for ``new physics'', it does merit watching as low $Q^2$ measurements of $\sstwQ$ along with
independent constraints on $Z_d$ mixing improve.

We start our discussion of intermediate mass $Z_d$ by briefly recalling its basic features.
That scenario assumes
a $U(1)_d$ gauge symmetry associated with a hidden dark sector.  Its gauge boson, $Z_d$, couples to our world (SM)
via kinetic mixing, parametrized by $\eps$, and $Z$-$Z_d$ mass matrix mixing, parametrized by $\eps_Z = (\mzd/m_Z)\delta$ \cite{Davoudiasl:2012ag}
\footnote{
We note that a new Higgs doublet charged under $U(1)_d$, assumed in typical models of $Z$-$Z_d$ mass mixing discussed in Ref.~[33], can also lead to non-zero kinetic mixing, via loop effects.}.
Actually, for an intermediate mass $Z_d$, the combination
\beq
\delta' \simeq \delta + \frac{\mzd}{m_Z} \,\eps \, \tan\theta_W
\label{delta'}
\eeq
proves important, as it governs the induced weak neutral current interactions of $Z_d$ (throughout our discussion, we ignore higher order corrections in $\eps$ and $\delta$).
It means the $\delta$ is replaced by the more general $\delta'$ of \eq{delta'} for an intermediate mass $Z_d$.
For the usually considered case of $\mzd\ll m_Z$, the second term in \eq{delta'} \cite{Gopalakrishna:2008dv} 
is generally negligible and $\delta' \simeq \delta$
becomes a good approximation, but here it is retained.
Depending on the relative sign of $\delta$ and $\eps$, the $Z$-$Z_d$ mass mixing or $\delta'$ might increase or decrease as $m_{Z_d}$ increases.

As a result of mixing, $Z_d$ couples to the SM via \cite{Davoudiasl:2012ag}
\beq
{\cal L}_{\text{int}} = (- e \eps J^{em }_\mu - \frac{g}{2 \cos \theta_W}\frac{\mzd}{m_Z} \delta' J^{NC}_\mu + \ldots ) Z_d^\mu,
\label{Lint}
\eeq
where the ellipsis represents other induced $Z_d$ interactions such as the $HZZ_d$ coupling \cite{Davoudiasl:2012ag,Davoudiasl:2012ig,Davoudiasl:2013aya} that we subsequently employ.
As a consequence of \eq{Lint}, weak neutral current SM amplitudes at low $Q^2$ momentum transfer are rescaled by $\rho_d$ (that is $\rho_d G_F$ instead of $G_F$) and the SM weak mixing angle $\sin^2\theta_W(Q^2)_{\text{SM}}$ is replaced by $\kappa_d \, \sin^2\theta_W(Q^2)_{\text{SM}}$ \cite{Davoudiasl:2012ag,Davoudiasl:2012qa,Davoudiasl:2014kua}
with
\beq
\rho_d = 1 + \delta'^2 \frac{\mzd^2}{Q^2 + \mzd^2}
\label{rhod}
\eeq
and
\beq
\kappa_d = 1 - \eps \delta' \frac{m_Z}{\mzd} \cot\theta_W \frac{\mzd^2}{Q^2 + \mzd^2}.
\label{kappad}
\eeq
The above yields a low $Q^2\ll m_{Z_d}^2$ shift
\bea
\Delta \sin^2\theta_W &\simeq& -\eps \delta' \frac{m_Z}{\mzd} \cos \theta_W \sin \theta_W \nonumber \\
&\simeq& -0.42 \,\eps \delta' \,\frac{m_Z}{\mzd}.
\label{Delsin2}
\eea

Note that the effect of $\rho_d$  in \eq{rhod} on $\sstwQ$ is process dependent.
Its largest effect is on the NuTeV result of \eq{sin2-NuTeV}, where an upward shift in the experimental $\sstwmZ$ of $\delta'^2$ 
is induced if $R_\nu$ (the ratio of neutral current to charged current neutrino cross sections) is employed \cite{Marciano:1980pb,Sirlin:1981yz}, and  
$\delta'^2/2$ if the Paschos-Wolfenstein relation \cite{Paschos:1972kj} is used.  Overall, $\rho_d$ has little effect 
on the weighted average in \eq{sin2-lowQ2}.
Nevertheless, including the effect of $\rho_d$ in future more precise studies is warranted.

As can be seen from \eq{Delsin2}, the value of $\sstwQ$ in our framework depends on 
$\mzd$, $\eps$, and $\delta'$.  Let us then consider next the current constraints on the latter two 
quantities over the $\mzd$ range of interest here.

\begin{figure}[t]
\includegraphics[width=0.48\textwidth]{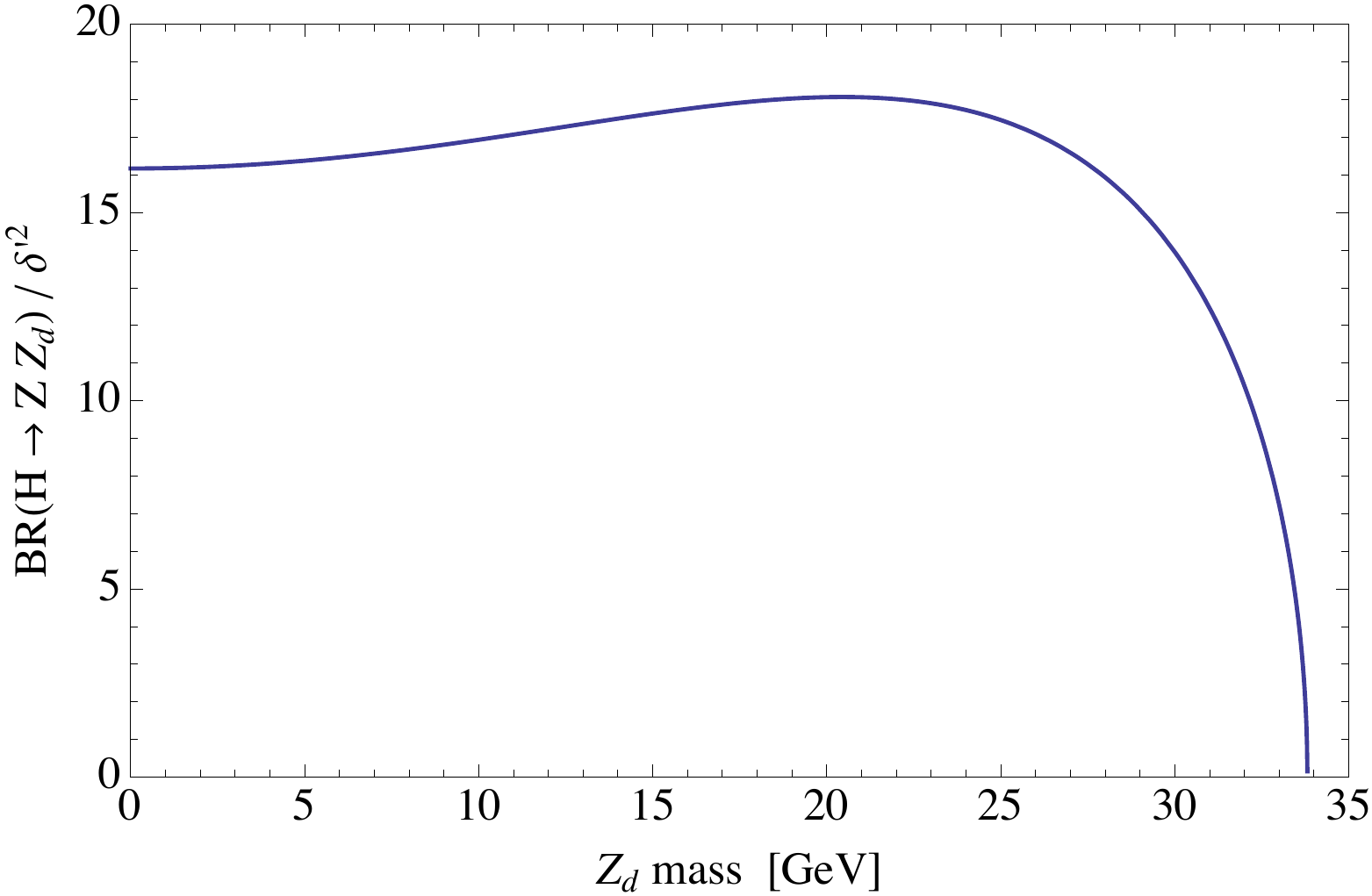}
\caption{$\br(H \to Z Z_d) / \delta'^2$ with $m_{Z_d}$. For the most part ($m_{Z_d} \lsim 30 ~\gev$), the branching ratio into $Z Z_d$ is almost independent of $m_{Z_d}$. $\br(H \to Z Z_d) \approx (16-18) \, \delta'^2$.}
\label{fig:BRHtoZZd}
\end{figure}
\begin{figure*}[t]
\begin{center}
\subfigure[]{
\includegraphics[width=0.48\textwidth,clip]{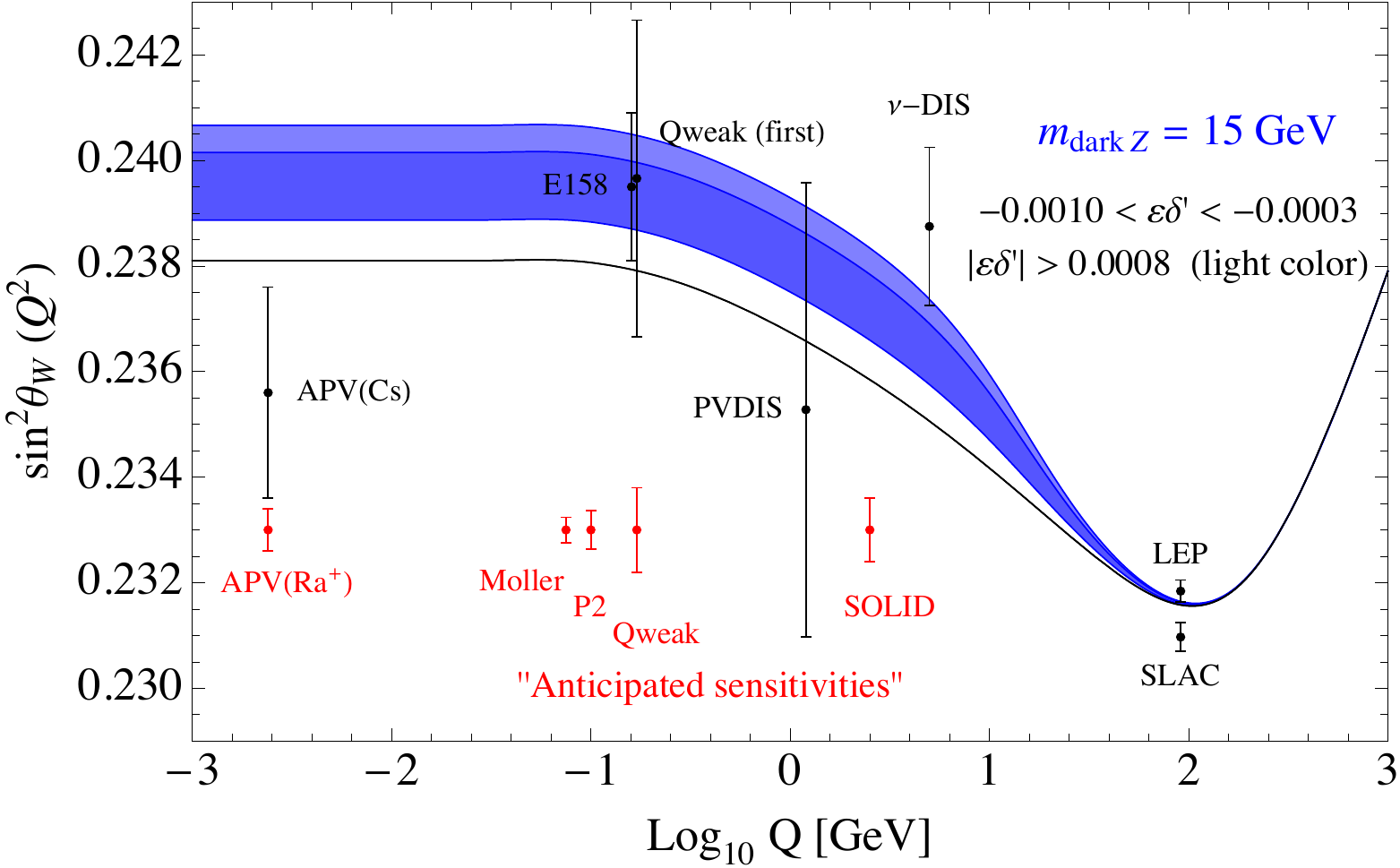}}
\subfigure[]{
\includegraphics[width=0.48\textwidth,clip]{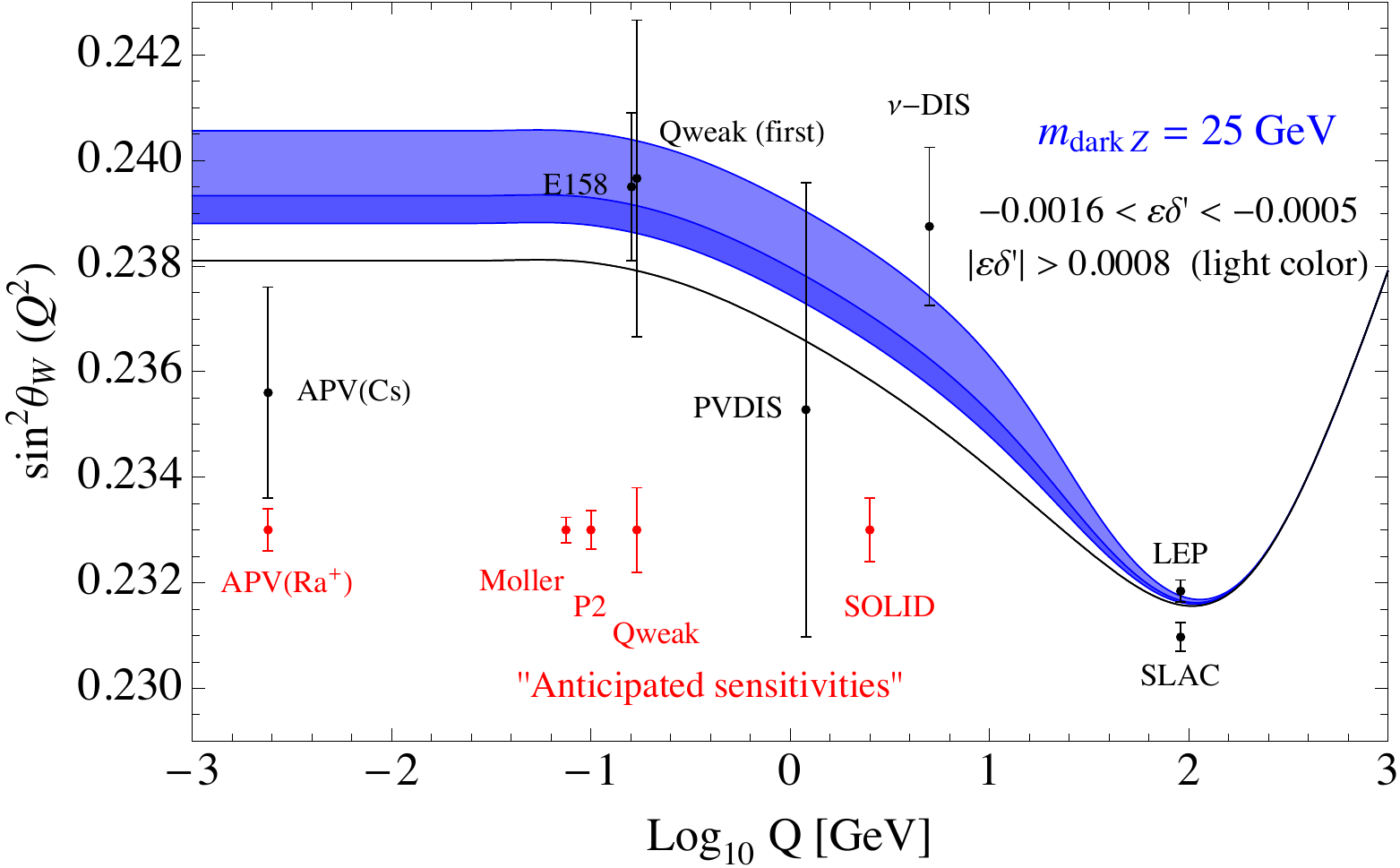}}
\end{center}
\caption{
Effective weak mixing angle running as a function of $Q^2$ shift (the blue band) due to an intermediate mass $Z_d$ for (a) $m_{Z_d} = 15$ GeV and (b) $m_{Z_d} = 25$ GeV for 1 sigma fit to $\eps \delta'$ in \eq{Delsin2}.
The lightly shaded area in each band corresponds to choice of parameters that is in some tension with precision constraints (see text for more details).
}
\label{fig:ZdRun}
\end{figure*}

Recently, the ATLAS collaboration at the LHC has reported results for 
the rare Higgs decay $H\to Z Z_d \to \ell^+_1 \ell^-_1 \ell^+_2 \ell^-_2$, with 
$\ell_{1,2}=e, \mu$ \cite{ATLAS-HZZd}. Assuming $Z$-$Z_d$ mass 
mixing parametrized by $\delta'$ and a dominantly SM-like Higgs boson of 125 GeV, 
one can show \cite{Davoudiasl:2012ag} that this decay has a branching ratio (roughly including $Z_d$ phase space effects \cite{Davoudiasl:2013aya})
\beq
\br(H\to Z Z_d) \approx (16-18) \, \delta'^2
\label{BrHZZd}
\eeq
which is further reduced by $Z$ and $Z_d$ leptonic branching ratios.
The on-shell branching ratio is given by \cite{Davoudiasl:2012ag,Davoudiasl:2013aya}
\bea
\br (H \to Z Z_d) = \frac{1}{\Gamma_H} \frac{\sqrt{\lambda(m_H^2, m_Z^2, m_{Z_d}^2)}}{16 \pi \, m_H^3} \left(\frac{g \, m_Z}{\cos\theta_W}\right)^2 \nn \\
\times \left( \delta' \frac{m_{Z_d}}{m_Z} \right)^2 \left( \frac{(m_H^2 - m_Z^2 - m_{Z_d}^2)^2}{4 m_Z^2 m_{Z_d}^2} + 2 \right) ~~~~~~~
\eea
with $\lambda(x, y, z) \equiv x^2 + y^2 + z^2 - 2xy - 2yz - 2zx$ and $\Gamma_H (125 ~\gev) \simeq 4.1 ~\mev$ \cite{Dittmaier:2012vm}, which shows a rather $m_{Z_d}$ independent value over most of the mass range (Fig.~\ref{fig:BRHtoZZd}), resulting in \eq{BrHZZd}.

The ATLAS bounds translate into constraints on $\delta'$ as a function of $\mzd$, but depend 
on the branching ratio for $Z_d\to \ell^+ \ell^-$.
For $\br(Z_d \to 2\ell) \equiv \br(Z_d \to 2e) + \br(Z_d \to 2\mu) \approx 0.3$ \cite{Batell:2009yf}, 
one finds (at 2 sigma) the nearly constant bound $|\delta'| \lsim 0.02$, over the range of $\mzd$ considered in our work.   
Here we note that in the presence of allowed dark decay channels (that is, decay into invisible particles),
$\br(Z_d \to 2\ell)$ can be much smaller than $ 0.3$,
which would weaken the constraint on $\delta'$.

The best current bounds on $\eps$ for the relevant mass range are given by the precision electroweak constraints, along with the non-continuous bounds from the $e^+ e^- \to$ hadron cross-section measurements at various experiments \cite{Hook:2010tw}.  
The Drell-Yan dilepton resonance searches at the LHC experiments (such as in Refs.~\cite{Chatrchyan:2012am,Chatrchyan:2013tia}) have the potential to give a better bound than precision electroweak constraints \cite{Hoenig:2014dsa}.  When combined with bounds on $\eps$ from precision measurements and production constraints \cite{Hook:2010tw,Curtin:2014cca}, 
one finds $|\eps| \lsim 0.03$, for kinetic mixing alone.  However, in our scenario, where a 
separate source of mass mixing is also considered \cite{Davoudiasl:2012ag}, that bound 
can be somewhat relaxed, via partial cancellation with $\delta'$ dependent contributions to the $Z$-$Z_d$ mixing angle \cite{Davoudiasl:2012ag}, roughly yielding $|\eps| \lsim 0.04$.   
(See also Refs.~\cite{Curtin:2013fra,Curtin:2014cca} for less severe bounds on $\eps$ from a recasting of a 
CMS analysis of Run 1 data, sensitive to $H\to Z Z_d$.)

Given the above discussion, a simple combination of the upper bounds on $\eps$ and $\delta'$ suggests 
\beq
|\eps \delta'| \lsim 0.0008.
\label{epsdelta'}
\eeq
We use the above bound as a rough guide for the allowed region of parameter space 
in our discussion below.  

For a given $\mzd$, a negative $\eps \delta'$ in \eq{Delsin2}
will shift the SM prediction in \eq{sin2-SM} towards the low $Q^2$
experimental $\sstwmZ$ weighted average in \eq{sin2-lowQ2}.  
That effect is illustrated in Fig.~\ref{fig:ZdRun} (a), where for $m_{Z_d}=15$~GeV the blue band corresponds to a 1-$\sigma$ fit to \eq{sin2-dev} or $-0.0010 < \eps \delta' < -0.0003$.
A similar 1-$\sigma$ band is presented in  Fig.~\ref{fig:ZdRun} (b) for $m_{Z_d}=25$~GeV with $-0.0016 < \eps \delta' < -0.0005$.
In each case, the lighter shaded upper part of the band corresponds to $|\eps \delta'|>0.0008$ which is in some tension with constraints from precision measurements and the rare Higgs decay search by ATLAS, as explained above.  
Future improved sensitivity at the LHC should cover most of the bands in Figs.~\ref{fig:ZdRun} (a) and (b).
For other $m_{Z_d}$ values, the 1-$\sigma$ bands are about the same as our Fig.~\ref{fig:ZdRun} representative examples; however, for larger $m_{Z_d} > 25 ~\gev$, the darker parts of the bands allowed by current constraints narrow.  
This can be seen from a comparison of Figs. 3 (a) and (b) that shows how smaller values of $\mzd$ can accommodate a shift in $\sstwQ$ more easily, over the currently allowed parameter space [as suggested by the $\mzd$ dependence in Eq.(12)]. 

In the case of low $Q^2$ determinations of $\sstwQ$, the Qweak polarized 
$e\,p$ asymmetry experiment at JLAB, which measures weak nuclear charge of proton ($Q_\text{weak}^p$), is expected to reach an uncertainty of $\pm 0.0007$ after all existing data 
are analyzed in the near future.  This would reduce the uncertainty 
on the weighted average in \eq{sin2-lowQ2} to $\pm 0.00055$ and, assuming the same central value 
as the current published result, could yield a $\sim 3$ $\sigma$ deviation from the SM result in 
\eq{sin2-SM}.  It will be interesting to watch that outcome.
We note that the weak mixing angle extracted from the Qweak experiment will exhibit some dependence on nucleon form factors including strangeness matrix element effects \cite{GonzalezJimenez:2011fq,Gonzalez-Jimenez:2014bia}.
For that reason, lattice gauge theory improvements in those hadronic matrix elements are strongly warranted.

Future experiments, primarily polarized $e \, e$ Moller scattering at JLAB and polarized $e\, p$ scattering (P2) at MESA in Mainz, are expected collectively to further reduce the weighted average uncertainty on $\sstwmZ$ at low $Q^2$
below $\pm 0.0002$, becoming competitive with $Z$ pole measurements.  
Together, low $Q^2$ precision studies combined with improved $H\to Z Z_d$ searches 
at the LHC will squeeze the intermediate mass $Z_d$ scenario with some possibility of uncovering 
its existence.

The intermediate mass $Z_d$ is an interesting viable alternative to 
the ``light" dark photon often considered in the literature \cite{Essig:2013lka}.  In addition to the parity violation at low $Q^2$ 
that we have explored, it can give rise to potential signals at the LHC, both in direct Drell-Yan production $p\,p \to Z_d\,X$ or as a final state in rare Higgs decays.  Besides the $H \to Z\, Z_d$ mode that we have discussed, searching for the mode $H \to Z_d\, Z_d$, mediated by 
Higgs-dark Higgs mixing \cite{Gopalakrishna:2008dv}, is well motivated.  In fact, we note that the ATLAS 8 TeV search for $H \to Z_d\, Z_d$ has two interesting but tentative candidate events (each at 1.7 $\sigma$), roughly in the mass range $\sim 20-25$~GeV \cite{ATLAS-HZZd}.   Further data from Run 2 at the LHC will be needed to clarify whether these events could be identified as intermediate mass $Z_d$ states that connect our world to an as yet unknown dark sector of Nature.   
Such a discovery would certainly revolutionize elementary particle physics and perhaps provide a new window into the world of dark matter.

\vspace{5mm}
Acknowledgments:
We thank Ketevi Assamagan and Keith Baker for discussions concerning the ATLAS dark vector boson searches.
The work of H.D. and W.J.M. is supported in part by the United
States Department of Energy under Award No. DESC0012704.
W.J.M. acknowledges partial support as a Fellow in the Gutenberg Research College.
The work of H.L. is supported in part by the CERN-Korea fellowship.



\end{document}